\documentclass[12pt]{iopart}
\usepackage{bm}
\usepackage{graphics}
\usepackage[dvipdfm]{graphicx}
\usepackage[dvips]{color}
\usepackage{dcolumn}

\begin{document}
\title{Instability of a superfluid Bose gas induced by a locked thermal gas in an optical lattice}
\author{S. Konabe and T. Nikuni}
\address{Department of Physics, Faculty of Science, Tokyo University of Science, 
1-3 Kagurazaka, Shinjuku-ku, Tokyo, Japan, 162-8601}
\ead{a1203622@rs.kagu.tus.ac.jp}

\begin{abstract}
We use a dissipative Gross-Petaevskii equation derived from the Bose-Hubbard Hamiltonian to study the effect of the thermal component on the stability of a current-carrying superfluid state of a Bose gas in an optical lattice potential. 
We explicitly show that the superfluid state becomes unstable at certain quasi-momentum of the condensate due to a thermal component, which is locked by an optical lattice potential. It is shown that this instability coincides with the Landau instability derived from the GP equation.
\end{abstract}


\maketitle
\section{Introduction}
Recently neutral atomic gases in an optical lattice potential have attracted much interest in the condensed matter people. 
This is because an optical lattice potential is the ideal realization of the solid state physics such as the formation of energy bands, Bloch oscillations, and Josephson effects, and parameters of the system can be easily tuned experimentally. 
More exciting feature appears in the combination of a Bose condensate and optical lattice potentials. These systems show the macroscopic quantum phenomena due to the phase coherence of the Bose-Einstein condensate such as the Mott insulator-superfluid transition realized in recent experiments~\cite{greiner2002,xu2005}.

In particular, properties of collective modes of a Bose condensed gas in an optical lattice are extensively studied both experimentally~\cite{cataliotti2001,burger2001,ferlaino2002,fallani2004,fertig2005,sarlo2005} and theoretically~\cite{wu2001,wu2003,smerzi2002,menotti2003,modugno2004,polkovnikov2004,altman2005} in the last few years.  
Much of attention is attracted to the phenomena of instability of a superfluid state.
Within the Gross-Pitaevskii (GP) theory in a periodic potential, two different mechanisms of the instability are known~\cite{wu2001,smerzi2002,wu2003,machholm2003}. 
One is  Landau or energetic instability which occurs when the excitation energy becomes negative. This kind of the instability is general in the superfluidity, not restricted in the lattice system. The other is dynamical or modulational instability which occurs when the excitation energy becomes imaginary. Such an instability typically exists in non-linear equations. 
It is important to note that although the dynamical instability occurs at zero temperature and can be studied using the GP equation, the Landau instability generally occurs at finite temperature. Therefore one cannot study the microscopic process of the Landau instability by using the GP theory. 
 
In experiments at finite temperatures such as Refs.~\cite{burger2001,ferlaino2002,sarlo2005}, there exists a thermal component around the condensate. As mentioned above, these situations cannot be investigated by the usual GP equation since it does not include a thermal component. In recent literature about the theory of a Bose gas in an optical lattice, however, detailed work considering a role of non-condensed atoms in the instability of the superfluidity is lacking. Thus, it is important to include the effect of a thermal component into the theory to go beyond the GP theory and give a microscopic physical picture of the Landau instability. 
In the present paper, we use a dissipative GP equation, which includes the effect of a thermal component, and explicitly show that non-condensed atoms locked by an optical lattice potential induce an instability of a current-carrying superfluid state. Our argument clarifies the physical picture of the Landau instability of a Bose gas in an optical lattice, the physical process of which is usually ambiguous in general discussions.    

The present paper is organized as follows. In Sec.~\ref{sec:dissipative}, we give a dissipative GP equation derived from the Bose-Hubbard Hamiltonian in an optical lattice potential, which includes a new term expressing collisions between a condensed and non-condensed atom. 
From the dissipative GP equation, we derive damped Bogoliubov equations in which an imaginary term due to a thermal component exists and then solve these equations to obtain the excitation energies. This procedure is done for a current-free state in Sec.~\ref{subsec:damped_damping} and a current-carrying state in Sec.~\ref{subsec:damped_instability}. We show that due to collisions between condensed and non-condensed atoms, the damping~(Sec.\ref{subsec:damped_damping}) and instability~(Sec.\ref{subsec:damped_instability})  appear for a current-free and current-carrying state, respectively.
In Sec.~\ref{subsec:damped_instability}, we also discuss the instability obtained in the present paper and the relation to the previous works about the instability of the usual GP theory. 
In Sec.~\ref{sec:summary}, we summarize our results.
 
\section{A dissipative Gross-Pitaevskii equation in an optical lattice}\label{sec:dissipative}
In this section, we derive an equation of motion for a Bose condensate order parameter which include the effect of a noncondensed atoms.

As a model of Bose atoms in a one-dimensional optical lattice, we use the following Bose-Hubbard Hamiltonian :
\begin{eqnarray}
\mathcal{H}=-J\sum_{i}\left(b_i^{\dag}b_{i+1}+b_{i+1}^{\dag}b_{i}\right)+\frac{U}{2}\sum_{i}b_i^{\dag}b_i^{\dag}b_ib_i,\label{bose-hubbard}
\end{eqnarray}
where $b_i$ and $b_i^{\dag}$ are annihilation and creation operators for Bose atoms at lattice site $i$, respectively. 
$J$ and $U$ are hopping parameter and onsite interaction, respectively. 
From the Bose-Hubbard Hamiltonian (\ref{bose-hubbard}), one can derive the dissipative GP equation~\cite{zaremba1999,imamovic1999,imamovic2000,stoof1999}
\begin{eqnarray}
\fl i\hbar\frac{\partial}{\partial t}\Psi_i(t)=-J\left[\Psi_{i+1}(t)+\Psi_{i-1}\right]+
\left\{U\left[n_i^c(t)+2\tilde{n}_i(t)\right]-i\hbar R({\bf r}_it)\right\}\Psi_i(t),\label{dissipative_GP}
\end{eqnarray}
where $\Psi_i\equiv\langle b_i\rangle$ is an order parameter of a condensate. 
In Eq.(\ref{dissipative_GP}), we defined the dissipative term $R$ as
\begin{eqnarray}
\fl R_i(t)\equiv\frac{\pi}{\hbar}\left(\frac{U}{2 N_{\rm site}}\right)^2\sum_{{k}_1,{k}_2,{k}_3}\delta\left({k}_c-{k}_1-{k}_2+{k}_3\right)\nonumber\\
\times\delta\biggl(\epsilon_i^c(t)-E_i({k}_1,t)-E_i({k}_2,t)+E_i({k}_3,t)\biggl)\nonumber\\
\times\biggl[
\left\{1+f_i({k}_1,t)\right\}\left\{1+f_i({k}_2,t)\right\}f_i({k}_3,t)\nonumber\\
\quad
-f_i({k}_1,t)f_i({k}_2,t)\left\{1+f_i({k}_3,t)\right\}\biggl].\label{dissipative_term}
\end{eqnarray}
The condensate momentum $k_c$ and the condensate energy $\epsilon_i^c(t)$ will be defined later.
The noncondensate atoms are treated in the Hartree-Fock (HF) approximation :
\begin{eqnarray}
E_i({k}_j)
&\equiv&-\epsilon({k}_j)+2U\left\{n_i^c(t)+\tilde{n}_i(t)\right\},\quad(j=1,2,3),\label{hartree-fock}
\end{eqnarray}
where $\epsilon(k_j)\equiv 2J\cos(k_j a)$ with $a$ being a lattice constant.
The dissipative term given by Eq.~(\ref{dissipative_term}) describes collisions between the condensate and the noncondensate atoms~\cite{zaremba1999}. 
Here, we used the following two approximations in deriving the dissipative GP equation~(\ref{dissipative_GP}).
First, we use the Hartree-Fock approximation for the excitation energies of the noncondensed atoms. This Hartree-Fock approximation used in Eq.(\ref{hartree-fock}) is valid only when $Un^c/J\ll1$. 
In this regime, there is no difference between the Bogoliubov and Hartree-Fock approximation~\cite{tsuchiya2004}.
Second, we assume the non-condensed atoms occupies only the first band. This assumption is valid when the band-gap between the first and second band is large compared to the thermal energy, i.e., $2k_BT\ll sE_R$, where $s$ is the strength of the optical lattice potential and $E_R$ is the recoil energy.

The dissipative term $R_i(t)$ is related to the collision term $C_{12}$ in the Boltzmann equation for the noncondensate atoms~\cite{zaremba1999} which describes the dynamics of  a nonequilibrium distribution function.
In the present study, we consider the situation where the thermal cloud is locked by an optical lattice potential and thus we assume that the thermal cloud is sufficiently close to static equilibrium. Instead of a nonequilibrium distribution function, we use the Bose distribution function
\begin{eqnarray}
f_i({k},t)
&\simeq& f^0({k})\nonumber\\
&=&\frac{1}{\exp\left\{\beta\left[-\epsilon({k})+2U\left(n^{c0}+\tilde{n}^0\right)-\tilde{\mu}_0\right]\right\}-1},\label{distribution_function}
\end{eqnarray}
where $n^{c0}$ and $\tilde{n}^0$ are the equilibrium densities of the condensate and the noncondensate, respectively and $\tilde{\mu}_0$ is a chemical potential of the non-condensed atoms in equilibrium. 
Denoting the dissipative term in which $f$ is displaced by $f^0$ in Eq.~(\ref{dissipative_term}) as $R^0({\bf r}_it)$ , the dissipative GP equation (\ref{dissipative_GP}) becomes
\begin{eqnarray}
\fl i\hbar\frac{\partial}{\partial t}\Psi_i(t)=-J\left[\Psi_{i+1}(t)+\Psi_{i-1}(t)\right]
\left\{U\left[n_i^c(t)+2\tilde{n}_i\right]-i\hbar R_i^0(t)\right\}\Psi_i(t).\label{dissipative_GP_2}
\end{eqnarray}

Using the equilibrium Bose distribution~(\ref{distribution_function}), one finds that the dissipative term is given by~\cite{williams2001_1, williams2001_2,zaremba1999}
\begin{eqnarray}
R_i^0(t)=\frac{1}{\tau_i(t)}\left\{e^{-\beta\left[\tilde{\mu}-\epsilon_i^c(t)\right]}-1\right\},
\end{eqnarray}
where we defined the collision time
\begin{eqnarray}
\fl \frac{1}{\tau_i(t)}
\equiv \frac{1}{\hbar}\left(\frac{U}{2 N_{\rm site}}\right)^2\sum_{{k}_1,{k}_2,{k}_3}\delta\left({k}_c-{k}_1-{k}_2+{k}_3\right)\nonumber\\
\times \delta\left(\epsilon_i^c(t)-E_i({k}_1,t)-E_i({k}_2,t)+E_i({k}_3,t)\right)\nonumber\\
\times f^0({k}_1)f^0({k}_2)\left[1+f^0({k}_3)\right].\label{collision_time}
\end{eqnarray}

Writing the order parameter $\Psi_i(t)$ in terms of the amplitude and phase,
\begin{eqnarray}
\Psi_i(t)=\sqrt{n_i^c(t)}e^{i\theta_i(t)}, 
\end{eqnarray}
we rewrite the dissipative GP equation (\ref{dissipative_GP}) as
\begin{eqnarray}
\fl\hbar\frac{\partial n_i^c(t)}{\partial t}
=-2J\sqrt{n_i^c(t)n_{i+1}^c(t)}\sin\left(\theta_{i+1}(t)-\theta_i(t)\right)\nonumber\\
+2J\sqrt{n_i^c(t)n_{i-1}^c(t)}\sin\left(\theta_i(t)-\theta_{i-1}(t)\right)+\hbar n_i^cR^0,\label{amplitude}
\end{eqnarray}
\begin{eqnarray}
\fl\hbar\frac{\partial \theta_i(t)}{\partial t}
=J\left[\sqrt{\frac{n_{i+1}^c(t)}{n_i^c(t)}}\cos\left(\theta_{i+1}(t)-\theta_i(t)\right)+\sqrt{\frac{n_{i-1}^c(t)}{n_i^c(t)}}\cos\left(\theta_i(t)-\theta_{i-1}(t)\right)\right]\nonumber\\
-U(n_i^c(t)+2\tilde{n}_i^0).\label{phase}
\end{eqnarray}
The right hand side of Eq.~(\ref{phase}) gives the energy of the condensed gas, i.e., 
\begin{eqnarray}
\fl\epsilon_i^c(t)
\equiv-J\left[\sqrt{\frac{n_{i+1}^c(t)}{n_i^c(t)}}\cos\left(\theta_{i+1}(t)-\theta_i(t)\right)+\sqrt{\frac{n_{i-1}^c(t)}{n_i^c(t)}}\cos\left(\theta_i(t)-\theta_{i-1}(t)\right)\right]\nonumber\\
-U(n_i^c(t)+2\tilde{n}_i^0).
\end{eqnarray}
In static thermal equilibrium, this energy of the condensate atom reduces to the chemical potential of the condensate $\mu_c^0$.

\section{Damped Bogoliubov equation for a current-free and current-carrying state}\label{sec:damped}
In this section, we derive Bogoliubov equations from the dissipative GP equation (\ref{dissipative_GP}) following Williams and Griffin~\cite{williams2001_2}. First, we derive Bogoliubov equations for a current-free state which include a damping term due to a thermal component, which we call damped Bogoliubov equations. Second, we generalize the damped Bogoliubov equations for a current-carrying state. 

\subsection{Damped Bogoliubov equation for a current-free state and damping of the excitation}\label{subsec:damped_damping}
In this section, we concentrate on the Bose condensate atoms at rest, i.e., $k_c=0$. If the gas is weakly disturbed from equilibrium, one can consider linearized solution by writing
\begin{eqnarray}
\Psi_i(t)=e^{-i\mu_c^0t/\hbar}\left[\Psi_0+\delta\Psi_i(t)\right],\label{ansaz1}
\end{eqnarray}
where $\mu^{c0}$ is a chemical potential of a condensate in static equilibrium.
Substituting Eq.(\ref{ansaz1}) into the dissipative GP equation (\ref{dissipative_GP_2}) and performing some algebra, one obtains the linearized equation of motion for a condensate fluctuation $\delta \Psi(t)$ :
\begin{eqnarray}
\fl i\hbar\frac{\partial}{\partial t}\delta\Psi_i(t)
=-J\left[\delta\Psi_{i+1}(t)+\delta\Psi_{i-1}(t)\right]\left[2U(n_i^{c0}+\tilde{n}_i^0)-\mu^{c0}\right]\delta\Psi_i(t)+Un^{c0}\delta\Psi_i^*(t)\nonumber\\
-i\frac{\hbar\beta n^{c0}}{\tau^0}\Biggl(
U\left[\delta\Psi_i(t)+\delta\Psi_i^*(t)\right]\nonumber\\
-\frac{J}{2n^{c0}}\biggl\{
\delta\Psi_{i+1}(t)+\delta\Psi_{i+1}^*(t)+\delta\Psi_{i-1}(t)+\delta\Psi_{i-1}^*(t)\nonumber\\
-2\left[\delta\Psi_i(t)+\delta\Psi_i^*(t)\right]\biggl\}\Biggl),\label{eq_of_motion_deltaPsi}
\end{eqnarray}
where we defined
\begin{eqnarray}
\fl\frac{1}{\tau^0}
\equiv\frac{1}{\hbar}\left(\frac{U}{2 N_{\rm site}}\right)^2\sum_{{k}_1,{k}_2,{k}_3}\delta\left(E({k}_2)+E({k}_3)-E({k}_1)\right)\nonumber\\
\times\delta\left({k}_2+{k}_3-{k}_1\right)f^0({k}_1)f^0({k}_2)\left\{1+f^0({k}_3)\right\}.
\end{eqnarray}

In order to solve Eq.(\ref{eq_of_motion_deltaPsi}), we expand $\delta\Psi_i$ as
\begin{eqnarray}
\delta\Psi_l(t)=\sum_q\left[u_qe^{i\left(qla/\hbar-\omega_qt\right)}+v_q^*e^{-i\left(qla/\hbar-\omega_q^*t\right)}\right],\label{fluctuation_current_free}
\end{eqnarray}
and substitute Eq.~(\ref{fluctuation_current_free}) into Eq.(\ref{eq_of_motion_deltaPsi}). We then obtain the damped Bogoliubov equations
\begin{eqnarray}
\hbar\omega_qu_q&=&\biggl[A(q)-i\frac{\hbar\beta}{2\tau^0} A(q)\biggl]u_q+\biggl[Un^{c0}-i\frac{\hbar\beta}{2\tau^0} A(q)\biggl]v_q,\label{bogoliubov_no_current1}\\
-\hbar\omega_qv_q&=&\biggl[A(q)+i\frac{\hbar\beta}{2\tau^0} A(q)\biggl]v_q+\biggl[Un^{c0}+i\frac{\hbar\beta}{2\tau^0} A(q)\biggl]u_q,\label{bogoliubov_no_current2}
\end{eqnarray}
where
\begin{eqnarray}
A(q)&\equiv&-2J\cos(qa)+2U\left(n^{c0}+\tilde{n}^0\right)-\mu^{c0}\nonumber\\
&=&4J\sin^2\left(\frac{qa}{2}\right)+Un^{c0}.
\end{eqnarray}
Solving the damped Bogoliubov equations (\ref{bogoliubov_no_current1}) and (\ref{bogoliubov_no_current2}) to first order in $1/\tau^0$, we find that the excitation energy is given by
\begin{eqnarray}
\fl\hbar\omega_{q}
=2\sin\left(\frac{qa}{2}\right)\sqrt{2UJn^{c0}+4J^2\sin^2\left(\frac{qa}{2}\right)}-i\frac{2\hbar\beta}{\tau^0}\biggl[2J\sin^2\left(\frac{qa}{2}\right)+Un^{c0}\biggl].\label{excitation_no_current}
\end{eqnarray}
One can see that the first term in this expression (\ref{excitation_no_current}) shows a usual Bogoliubov spectrum. For small $q$, the spectrum~(\ref{excitation_no_current}) reveals a phonon-like collective mode. 
On the other hand, the second term in Eq.(\ref{excitation_no_current}) works as the damping to the collective mode in a condensate since this term is always positive. This type of damping arises from the collisions between condensed and non-condensed atoms~\cite{williams2001_1,williams2001_2} and exists even if non-condensed atoms are in static equilibrium.
We note that there exists the Landau damping which originates from the interaction between a condensate collective mode and the excitations of the thermal component~\cite{tsuchiya2004}. To include the Landau damping, one has to consider the dynamics of noncondensed atoms.

\subsection{Damped Bogoliubov equation for a current-carrying state and instability of the superfluidity}\label{subsec:damped_instability}
In this section, we derive damped Bogoliubov equations for a current-carrying state, i.e., a condensate has a quasi-momentum $k_c\neq0$ and discuss the instability of the superfluidity.

Instead of Eq.(\ref{ansaz1}), we substitute the following ansatz 
\begin{eqnarray}
\Psi_l(t)=e^{i\left(k_cla/\hbar-\mu^{c0}t/\hbar\right)}\left[\Psi_0+\delta\Psi_l(t)\right],
\end{eqnarray}
into Eq.(\ref{dissipative_GP}).
Then one obtains an equation of motion for the fluctuation of the order parameter :
\begin{eqnarray}
\fl i\hbar\frac{\partial}{\partial t}\delta\Psi_i(t)
=-J\left[e^{ik_ca/\hbar}\delta\Psi_{i+1}(t)+e^{-ik_ca/\hbar}\delta\Psi_{i-1}(t)\right]+Un^{c0}\delta\Psi_i^*(t)\nonumber\\
+\left[2U(n^{c0}+\tilde{n}^0)-\mu_k^{c0}\right]\delta\Psi_i(t)-i\hbar\Psi_0\delta R^0(t)
,\label{eq_of_motion_deltaPsi_current}
\end{eqnarray}
where $\mu_k^{c0}\equiv-2J\cos(k_ca)+U(n^{c0}+2\tilde{n}^0)$ and 
\begin{eqnarray}
\fl\delta R^0(t)
=-\frac{\beta}{\tau^0}\mu_{\rm diff}(t)\nonumber\\
\!\!\!\!\!\!\!\!\!\!\!\!\!\!\!\!\!\!\!\!\!\!\!\!=-\frac{\beta}{\tau^0}\Biggl\{\frac{J}{2\sqrt{n^{c0}}}
\biggl[\delta\Psi_{l+1}(t)+\delta\Psi_{l+1}^*(t)+\delta\Psi_{l-1}(t)+\delta\Psi_{l-1}^*(t)\nonumber\\
-2\left(\delta\Psi_l(t)+\delta\Psi_l^*(t)\right)\biggl]\cos(k_ca)\nonumber\\
-\frac{J}{2i\sqrt{n^{c0}}}\left[\delta\Psi_{l+1}(t)-\delta\Psi_{l+1}^*(t)-\delta\Psi_{l-1}(t)+\delta\Psi_{l-1}^*(t)\right]\sin(k_ca)\nonumber\\
-U\sqrt{n^{c0}}\left[\delta\Psi_l(t)+\delta\Psi_l^*(t)\right]\Biggl\}.
\end{eqnarray}
Similarly to the current-free case, we substitute 
\begin{eqnarray}
\delta\Psi_l(t)=\sum_q\left[u_qe^{i\left(qla/\hbar-\omega_{kq}t\right)}+v_q^*e^{-i\left(qla/\hbar-\omega_{kq}^*t\right)}\right],
\end{eqnarray}
into Eq.(\ref{eq_of_motion_deltaPsi_current}).
We then obtain the equations similar to Eqs.(\ref{bogoliubov_no_current1}) and (\ref{bogoliubov_no_current2}) ;
\begin{eqnarray}
\fl\hbar\omega_{kq}u_q=\biggl[\tilde{A}(k_c+q)-i\frac{\hbar\beta}{2\tau^0} \tilde{A}(k_c-q)\biggl]u_q+\biggl[Un^{c0}-i\frac{\hbar\beta}{2\tau^0}\tilde{A}(k_c+q)\biggl]v_q,\label{bogoliubov_current1}
\end{eqnarray}
\begin{eqnarray}
\fl-\hbar\omega_{kq}v_q=\biggl[\tilde{A}(k_c-q)+i\frac{\hbar\beta}{2\tau^0} \tilde{A}(k_c+q)\biggl]v_q+\biggl[Un^{c0}+i\frac{\hbar\beta}{2\tau^0} \tilde{A}(k_c-q)\biggl]u_q,\label{bogoliubov_current2}
\end{eqnarray}
where
\begin{eqnarray}
\tilde{A}(k_c\pm q)&\equiv&Un^{c0}-2J\biggl[\cos\left((k_c\pm q)a\right)-\cos(k_ca)\biggl].
\end{eqnarray}

Using the damped Bogoliubov equations for the current-carrying state (\ref{bogoliubov_current1}) and (\ref{bogoliubov_current2}), one obtains excitation spectra.
First, we derive the excitation spectra from Eq.(\ref{bogoliubov_current1}) and (\ref{bogoliubov_current2}) when the imaginary terms are neglected. 
After some algebra, one obtains
\begin{eqnarray}
\fl \hbar\omega_{kq}=2J\sin(k_ca)\sin(qa)\nonumber\\
\pm 2\sin\left(\frac{qa}{2}\right)
\sqrt{2UJn^{c0}\cos(k_ca)+4J^2\cos^2(k_ca)\sin^2\left(\frac{qa}{2}\right)}.\label{excitation_without}
\end{eqnarray}
From this expression of the excitation spectra, one can see that the instability occurs when the excitation energy becomes negative, or when it develops the imaginary part, i.e., $k_c\ge \pi/2$.
One can identify the origin of the first instability as the Landau instability, and the second one as the dynamical instability, which were discussed in detail in Ref.~\cite{wu2001,wu2003}.

\begin{center}
\begin{figure}
    \begin{tabular}{cc}
      \scalebox{0.6}{\includegraphics{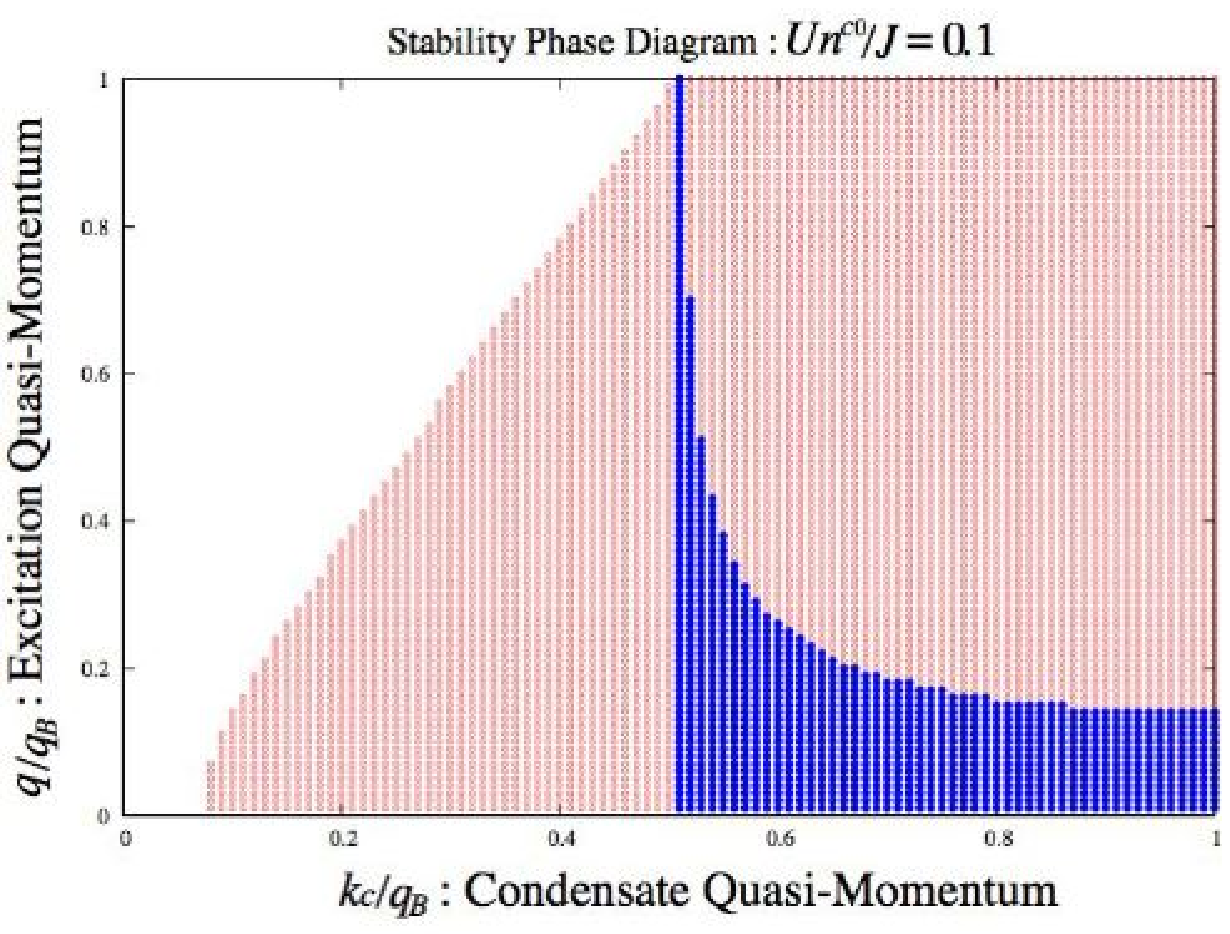}} &
      \scalebox{0.6}{\includegraphics{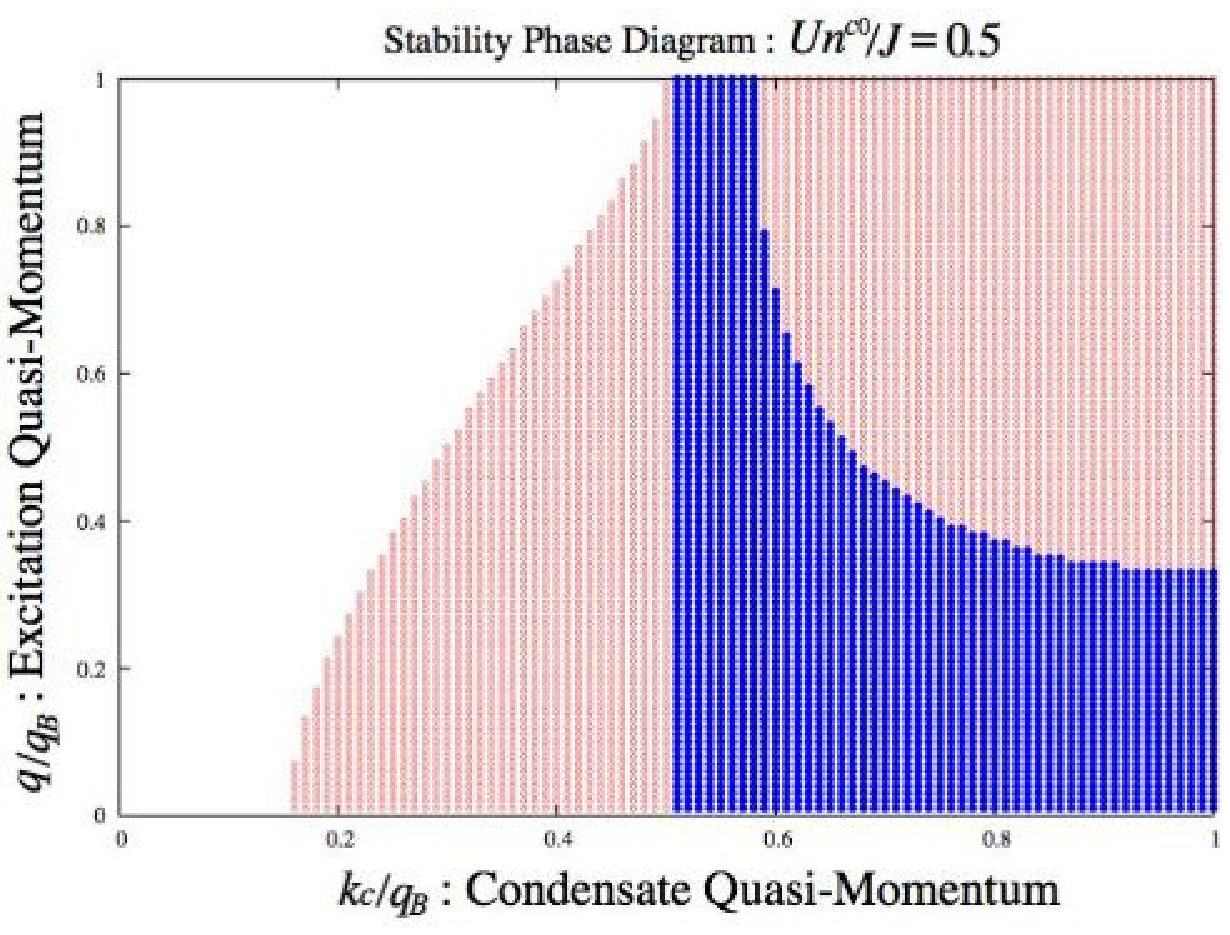}}
    \end{tabular}
    \caption{Stability phase diagram. The horizontal axis is the quasi-momentum of the condensate and the vertical axis is the quasi-momentum of the excitation. $q_B$ is the Bragg wave vector defined by $q_B=\pi/a$. Colored regions are unstable : Blue area is the dynamically unstable region and both red and blue areas are the Landau and dissipative unstable region.}
    \label{fig:stability}
\end{figure}
\end{center}

Second, we calculate the excitation spectra including the imaginary terms. We obtain the following expressions :
\begin{eqnarray}
\fl\hbar\omega_{kq}
=2J\sin(k_ca)\sin(qa)\nonumber\\
\pm 2\sin\left(\frac{qa}{2}\right)
\sqrt{2UJn^{c0}\cos(k_ca)+4J^2\cos^2(k_ca)\sin^2\left(\frac{qa}{2}\right)}\nonumber\\
-i\frac{2\hbar\beta}{\tau^0}\biggl[2J\cos(k_ca)\sin^2\left(\frac{qa}{2}\right)+J\frac{\alpha-1}{\alpha+1}\sin(k_ca)\sin(qa)+Un^{c0}\biggl],
\label{excitation_with}
\end{eqnarray}
where 
\begin{eqnarray}
\fl\alpha\equiv -1-\frac{4J}{Un^{c0}}\cos(k_ca)\sin^2\left(\frac{qa}{2}\right)\nonumber\\
+2\sin\left(\frac{qa}{2}\right)
\sqrt{\frac{2J}{Un^{c0}}\cos(k_ca)\left\{1+\frac{2J}{Un^{c0}}\cos(k_ca)\sin^2\left(\frac{qa}{2}\right)\right\}}.
\end{eqnarray}
Eq.(\ref{excitation_with})  is the main result of the present paper.
From Eq.(\ref{excitation_with}), one can find a new origin of the instability, which we call ``dissipative instability''.
When the sign of the third term of Eq.(\ref{excitation_with}) is changed at a certain quasi-momentum $k_c$, the excitation energy increases exponentially as time increases, which means that the superfluid state becomes unstable. This instability directly comes from the collisional term between the condensed and non-condensed atoms, and thus arises due to the mutual interaction between the condensed and noncondensed atoms.
The stable phase diagrams in $k_c$-$q$ plane are shown in Fig.\ref{fig:stability}. 
From Fig.\ref{fig:stability}, one can see that the critical line of the Landau instability, which is derived from the GP equation~\cite{wu2001,wu2003}, coincides with the critical line of the ``dissipative instability". 
This is physically reasonable because in experiments the thermal component should be playing a role of the obstacle that induces the Landau instability.
Here, we have explicitly shown that the collisions between the condensed and non-condensed atoms locked by an optical lattice potential indeed make the system unstable in the sense as the Landau instability. 
We note that this kind of instability is analogous to the instability of a condensate induced by a rotating thermal cloud discussed by Williams {\it et al.}~\cite{williams2002}.

\section{Summary}\label{sec:summary}
In the present paper, we have studied the instability of the current-carrying superfluid state at finite temperature in an optical lattice potential. Using the dissipative GP equation derived from the Bose-Hubbard Hamiltonian, we have obtained the damped Bogoliubov equations and calculated the excitation spectrum including the imaginary part due to a thermal component.
By analyzing the excitation spectrum, we have explicitly shown, for the first time, that  the mutual friction between the condensate and the thermal component locked by an optical lattice potential  leads to the instability at a critical condensate momentum, which is equivalent to the Landau instability criterion.   
This work gives the clear physical picture about the Landau instability of a Bose gas in an optical lattice at finite temperature in the actual experimental situation.

In the present work, we have only investigated the stable phase diagram focusing on the effect of the thermal component. It will be important and interesting to compare our analysis with the experimental data performed recently~\cite{sarlo2005}. 
In this connection, it will be interesting to investigate the temperature dependence of the imaginary term in the excitation spectrum and calculate the growing time of the excitations.  

\ack
We wish to thank Ippei Danshita for fruitful discussions. We also acknowledge Allan Griffin for useful comments. S.K. is supported by JSPS (Japan Society for the Promotion of Science) Research Fellowship for Young Scientists.

\end{document}